\documentclass[reprint,aps,prb,twocolumn,superscriptaddress]{revtex4-2}
\usepackage{graphicx}
\usepackage{mathrsfs}
\usepackage{mathtools}
\usepackage{bm}
\usepackage{hyperref}
\usepackage{dcolumn}
\usepackage{amsmath}
\usepackage{amssymb}    
\usepackage{color}
\usepackage{bbold}
\usepackage{braket}
\usepackage{xcolor}
\usepackage{ifpdf}

\newcommand{\bH}{{\mathbf H}}

\newcommand{\bn}{{\mathbf n}}
\newcommand{\bk}{{\mathbf k}}
\newcommand{\bj}{{\mathbf j}}
\newcommand{\bs}{{\mathbf \sigma}}

\usepackage[final]{changes}
\usepackage{braket}

\begin{document}
\title{Kerr, Faraday, and Magnetoelectric Effects in MnBi$_2$Te$_4$ Thin Films}
\author{Chao Lei}
\affiliation{Department of Physics, The University of Texas at Austin, Austin, Texas 78712, USA}
\email{leichao.ph@gmail.com}
\author{Allan H. MacDonald}
\affiliation{Department of Physics, The University of Texas at Austin, Austin, Texas 78712, USA}

\begin{abstract}
The topological magneto-electric effect (TME) is a characteristic property of 
topological insulators.  In this article, we use a simplified coupled-Dirac-cone 
electronic structure model to theoretically evaluate the THz and 
far infrared Kerr and Faraday responses of thin 
films of MnBi$_2$Te$_4$ with up to $N=10$ septuple layers with the goal of clarifying 
the relationship between these convenient magneto-optical observables and the TME.
We find that for even $N$ the linear Kerr and Faraday responses
to an electric field vanish in the low-frequency limit, 
even though the magnetoelectric response is large and approximately quantized.  
\end{abstract}

\maketitle

\section{Introduction}
Three-dimensional topological 
insulators (TIs) \cite{Hasan2010,Qi2011} have protected 
surface states with 
Dirac band-crossings located at time-reversal invariant two-dimensional momenta and characteristic topological magneto-electric (TME)
response\cite{Qi2008,Essin2009,Essin2010,Nomura2011,Wang2015,Morimoto2015}  properties.
The TME effect occurs only when the Dirac cones are gapped by introducing magnetic dopants at the 
surface or by using magnetic TIs like MnBi$_2$Te$_4$, and has been proven to be difficult to measure directly\cite{Fijalkowski2021,Chen_2023}. 
In the thin-film limit Kerr and Faraday's optical response coefficients
and orbital magnetization, all of which require broken time-reversal symmetry, are closely related quantities 
that are normally present or absent together.
Partly for this reason there has been interest \cite{Tse2011,Tse2010a,Tse2010b,Wu_Exp_2016,Dziom2017,Armitage_RMP_2018,Armitage_2019,he2022topological,Molenkamp_2022,Ahn_2022,Ghosh_2023}
in using magneto-optical Kerr or Faraday effects as a proxy for magnetization since Kerr and Faraday effect measurements are routinely 
used as a proxy for magnetization measurements.

The Kerr and Faraday effects of TIs are easily measured \cite{Fu2007,Brune2011,Kozlov2014,Xu_2014,Yoshimi_2015} 
when external fields are 
applied or the magnetization orientations on top and bottom surfaces are parallel, in which case the device Hall conductivity 
is quantized \cite{Yu2010,Liu2008,Chang2013} at a non-zero value and the magnetization is 
non-zero even in the absence of an electric field.  MnBi$_2$Te$_4$ films with an odd number of septuple layers $N$ in which the surface magnetizations are parallel provide one example of this quantum Hall case.
In this article, we exclude the quantum Hall devices from consideration and 
focus on the case of even $N$ magnetic topological insulators 
(and on even layer MnBi$_2$Te$_4$ in particular) instead of surface magnetized non-magnetic TIs 
since these seem at present to have more reproducible magnetic properties, although our conclusions apply to both cases. 

Our interest here is thus in the magnetization response to electric field in MnBi$_2$Te$_4$ films 
with an even number of septuple layers, in which
the TME coefficient is quantized and the total Hall conductivity in the absence of electric and magnetic field is zero.
We point out that in this case both the Kerr and the Faraday responses to an electric field 
differ qualitatively from the magnetization response. 
Specifically for even $N$ , the Kerr and Faraday angle response to an electric field vanishes at low frequencies
under circumstances where the TME is robust.
We explain this difference using a simplified coupled Dirac-cone model \cite{Lei2020} of magnetic TI MnBi$_2$Te$_4$.
Below we first explain the origin of this difference, and then explore it quantitatively using a simplified model of magnetic topological insulator thin films.

\section{Faraday, Kerr, and Magnetoelectric Response}

Hall effects in two-dimensional insulators can be viewed \cite{macdonald1995proceedings}
as measurements of the chemical potential dependence of equilibrium edge currents, $dI_{edge}/d\mu=\sigma e/h$,
or equivalently of the orbital magnetizations that they produce $dM_{edge}/d\mu = \sigma A e^2/h$,
where $e$ is the magnitude of the electron charge, $h$ is Plank's constant, $\sigma$ is an integer
and A is the film area.
In topological insulator thin films $\sigma$ 
can be non-zero only when time-reversal symmetry is broken,
either at the top and bottom surfaces or, as in the case of a magnetic topological insulator
\cite{Mong2010}, throughout the bulk.  In a system with A-type bulk antiferromagnetism,
the TME occurs when opposite surfaces have
opposite magnetizations, {\it e.g.} for an even number of magnetic layers.
The magnetic configuration with opposite magnetization orientations on opposite surfaces is often referred to in the literature as the {\it axion insulator} configuration.  
Since the edge current at a surface depends
only on the value of the chemical potential relative to the
mid-gap energies of the local Dirac cones, or some other reference energy,
it follows that $dM_{edge}/dV_{edge} = - dM_{edge}/d\mu$.  
When an electric field is applied across the bulk of the topological 
insulator with thickness $t$, the local electrical potentials on the top and bottom surfaces differ by $e \mathcal{E}_z t$, moving the 
local Dirac bands relative to the chemical potential as illustrated in Fig.~\ref{fig:schematic}.
This difference yields a net magnetization that is linear in $\mathcal{E}_z$ and proportional to the system volume - the topological magnetoelectric effect.  
The total Hall conductivity of the axion insulator state,
summing over top and bottom surfaces, still vanishes however provided that the chemical potential stays inside the surface state gap. 

\ifpdf
\begin{figure}[htb]
  \centering
  \includegraphics[width=0.9 \linewidth ]{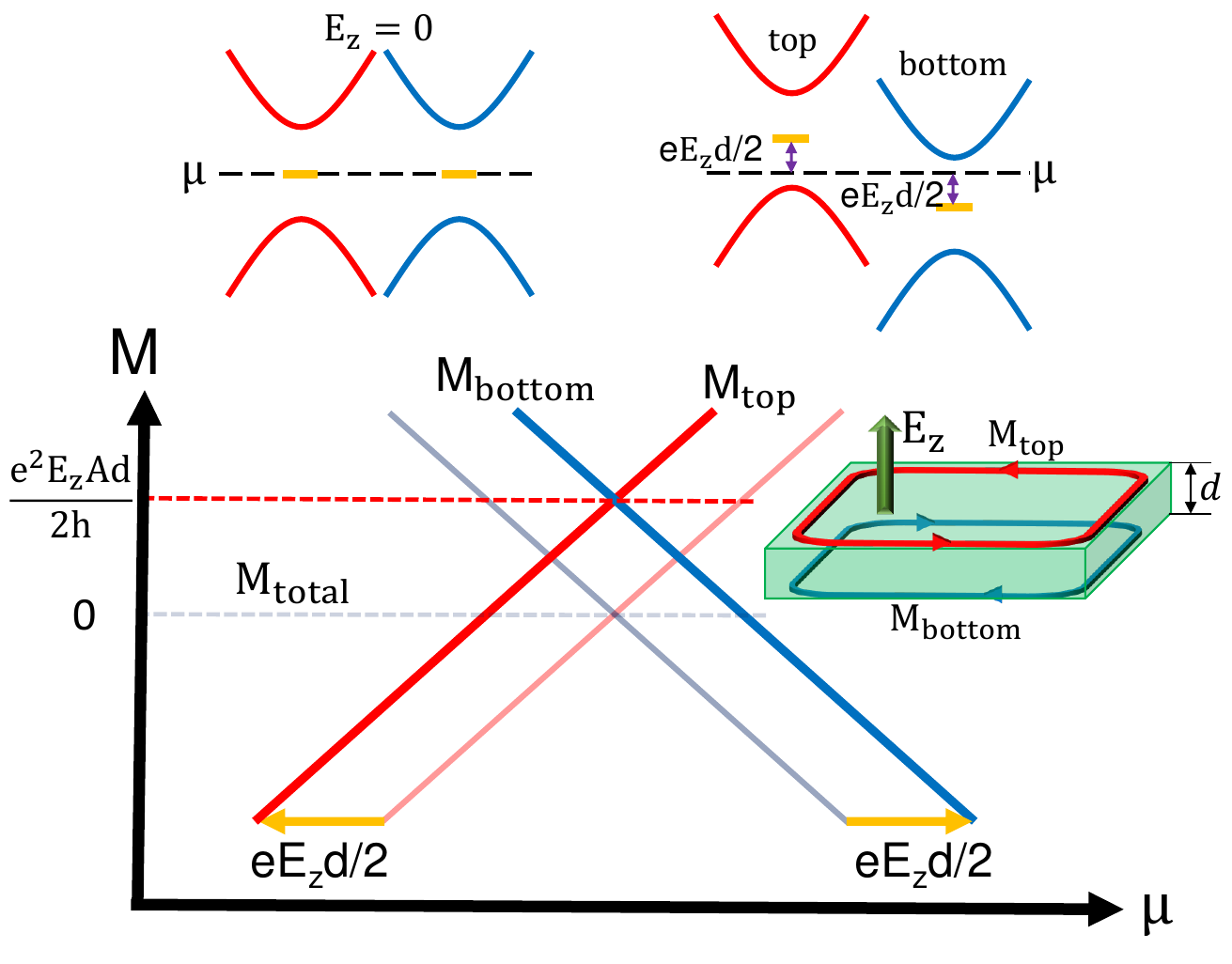}
  \caption{Magnetization due to current circulating around the top (red) and bottom (blue)
  surfaces of a magnetic TI.  At each surface, the magnetization depends linearly on the local chemical potential relative to the local electric potential.  
  When an electric field $\mathcal{E}_z$ is applied across the bulk of the thin film it produces a relative energetic shift between the gapped Dirac cones.  
  The total equilibrium orbital magnetization is then 
  proportional to $\mathcal{E}_z$ and the film volume.
    }\label{fig:schematic}
\end{figure}
\fi

The TME effect refers to the {\it dc} response properties of TIs, but is also approximately manifested at 
finite frequencies provided they are well below the TI bulk energy gap.
Since the gaps of MnBi$_2$Te$_4$ thin films are usually less than 100 meV, typical film 
thicknesses $d$ are very small compared to the relevant light wavelength $\lambda$.
(For example $d \sim 150$ nm for a $N=10$ septuple layer MnBi$_2$Te$_4$ thin film is small compared to the 
vacuum wavelength $\lambda \sim 25 \mu$m of $50$ meV light.) 
In the thin film ($d \ll \lambda$) limit
we can calculate the Kerr and Faraday responses simply by treating
the entire film as an arbitrarily thin two-dimensional interface.
The response of light to currents in the 2D film is determined
by the electromagnetic boundary condition,
\begin{equation}\label{eq:boundary}
    \bn \times (\bH_t - \bH_b) = \bj_s.
\end{equation}
Here $\bn$ is a unit vector oriented from top to bottom and $j_s$ is the two-dimensional current density,
which is related to the two-dimensional conductivity of the film by
\begin{equation}
    j_s^{\alpha} = \sum_{\beta} \sigma_{\alpha \beta} E_{\beta}.
\end{equation}
When Eq. \ref{eq:boundary} is combined with the source-free Maxwell's equations applied outside of the thin film,
the in-plane transmitted and reflected fields produced by an incident {\it em} wave with an unit electric field are \cite{Tse2011} 
\begin{equation}\label{eq:efield}
\begin{aligned}
& \begin{pmatrix}
E_x^{t}\\
E_y^{t}
\end{pmatrix} = \frac{1}{\mathcal{N}} 
    \begin{pmatrix}
2n_1(n_1+n_2+2\alpha \sigma_{xx})\\
-4 \alpha n_1 \sigma_{xy}
\end{pmatrix},\\
&\begin{pmatrix}
E_x^{r}\\
E_y^{r}
\end{pmatrix} = \frac{1}{\mathcal{N}} 
    \begin{pmatrix}
n_1^2 - (n_2+2\alpha \sigma_{xx})^2 - (2\alpha \sigma_{xy})^2 \\
-4 \alpha n_1 \sigma_{xy}
\end{pmatrix},
\end{aligned}
\end{equation}
where $\sigma_{xx}$ and $\sigma_{xy}$ are the total longitudinal and Hall conductivities of the film
in units of $e^2/h$, $\alpha \approx 1/137$ is the fine structure constant, 
$\mathcal{N} \equiv (n_1 + n_2 +2\alpha \sigma_{xx})^2 + (2\alpha \sigma_{xy})^2$, and 
$n_i^2 = \epsilon_i$ are the relative dielectric constants of the materials above and below the interface.  
The 2D film conductivities should be evaluated by integrating
across the film and include contributions from both dissipative 
and reactive responses of the TI film both at its surfaces and in the interior of the film.  
The 2D approximation, which has the advantage of allowing us to reach simple conclusions,
is strictly speaking valid only in the limit $d/\lambda \to 0$,
as discussed further below.

The Faraday and Kerr angles are defined respectively as the rotation angles of  
linearly polarized incoming light upon transmission and reflection:
\begin{equation}\label{eq:faraday_kerr}
 \begin{aligned}
  & \theta_F = (\arg{E_{+}^{t}} - \arg{E_{-}^{t}})/2,  \\
  & \theta_K = (\arg{E_{+}^{r}} - \arg{E_{-}^{r}})/2,  \\
 \end{aligned}
\end{equation}
where $E_{\pm}^{r/t} \equiv E_{x}^{r/t} \pm iE_{y}^{r/t}$.
Here the values of incoming in-plane polarization can be read from Eq.~\ref{eq:efield}.

For even $N$ films both the Hall conductivity and the magnetization vanish
by symmetry \cite{Lei2021_metamagnetism,lei2022quantum} in the absence of external out-of-plane electric field 
$\mathcal{E}_z=0$ at all frequencies.  Because of quantization, the {\em dc} Hall 
conductivity vanishes identically at finite $\mathcal{E}_z$ until the field is strong enough to close the gap; 
The linear response of the Hall conductivity to $\mathcal{E}_z$ is therefore zero in the {\em dc} limit.
Under the same conditions, the linear response of the magnetization is quantized at the topologically protected value.  
We anticipate that the linear response of the Kerr and Faraday effects to external electric field $E_z$
is strongly suppressed when $\hbar \omega$ is well below the bandgap of MnBi$_2$Te$_4$ thin films.  In the following we use a simplified model to test this expectation quantitatively. 

\section{Optical Conductivity of MBT \\ Thin Films}
We evaluate the frequency-dependent conductivity tensor of 
MnBi$_2$Te$_4$ \cite{Otrokov_2017,Eremeev2017,Otrokov2019,Otrokov2019_film,Rienks2019,Chen2019,Deng_2020,Ge2020,Li2019_theory,Chowdhury_2019,Li2020,Yan2019,Zeugner2019,Zhang2019_AHC,Liu2020,Lei2020} thin films
using a coupled-Dirac-cone model \cite{Lei2020} that retains two Dirac cones in 
each MnBi$_2$Te$_4$ septuple layer as low-energy degrees of freedom.  
The full Hamiltonian in the presence of external out-of-plane electric field reads as \cite{Lei2021_QAH}:
\begin{equation}
\label{eq:model}
\begin{split}
  H = & \sum_{\bk_{\perp},ij} \Big[\Big( \, 
  (-)^i  \hbar v_{_D}  (\hat{z} \times \bs) \cdot \bk_{\perp} + m_{i} \sigma_z + V_i \Big) \delta_{ij}   \\
  & + \Delta_{ij}(1-\delta_{ij} ) \Big] c_{\bk_{\perp} i}^{\dagger} c_{\bk_{\perp} j} ~.
\end{split}
\end{equation}
Here the Dirac cone labels $i$ and $j$ are respectively odd and even on the top and 
bottom surface of each septuple layer, $\hbar$ is the reduced Planck's constant, 
$v_{_D}$ is the Dirac-cone velocity and $V_i = V_{i-1} + \mathcal{E}_i  (z_i-z_{i-1})$ is the self-consistent Hartree potential on surface $i$, with $z_i$ the designed position of the $i^{th}$ Dirac cone.
The external electric field is calculated with discrete Poisson equation as $\tilde{\epsilon} \mathcal{E}_i = \tilde{\epsilon} \mathcal{E}_{i-1} + \delta \rho_i$, here $\tilde{\epsilon}$ is the dielectric constant and $\delta \rho_i$ are the net surface charge density at surface $i$, more details for the calculations of $\delta \rho_i$ and related parameters can be found in \onlinecite{Lei2021_QAH}. In the following discussion when the electric field is present, we will always consider the case when the Fermi level lies in the gap, i.e. we will keep the system neutral.
The Dirac-cone model describes a topological insulator when the 
hybridization $\Delta_{D}$ across the gap between different 
septuple layers is stronger than the hybridization $\Delta_{S}$ 
between top and bottom Dirac cones in the same septuple layer.  
Each Dirac cone has an exchange splitting $m$ that is the sum of 
contributions from the near-neighbor Mn magnetic layers within the 
same ($J_S$) septuple layer and in the adjacent septuple layer ($J_D)$.
An $N$-septuple layer thin film is reduced by this model to a quasi-2D system with 4N bands.  
For the calculations we report on below we use
the numerical model 
parameters that provide a minimal description\cite{Lei2020} of MnBi$_2$Te$_4$ thin films: 
Dirac velocity $\rm v_D = 5 \times 10^5 m/s $,
$\rm \Delta_S = 84 meV$, $\rm \Delta_D = -127 meV$, $\rm J_S = 36 meV$, and $\rm J_D = 29 meV$.

The optical conductivity of the Dirac cone model is calculated by using the Kubo-Greenwood 
formula\cite{Kubo1957,Greenwood_1958}:
\begin{equation}\label{eq:optical_conductivity}
\begin{split}
    \sigma_{\alpha \beta}(\omega) = & \frac{ie^2}{\hbar} \int \frac{d\bk}{(2\pi)^2} \sum_{nm} \frac{f_{n\bk} - f_{m\bk}}{E_{n\bk} - E_{m\bk}} \\
    & \times \frac{ \braket{m\bk | \partial_{\alpha} H_{\bk} | n\bk } \braket{n\bk | \partial_{\beta} H_{\bk} | m\bk } } {E_{n\bk} - E_{m\bk} -(\hbar \omega + i\eta)},
\end{split}
\end{equation}
where $\partial_{\alpha} H_{\bk} \equiv \partial_{k_\alpha} H_{\bk}$, $\omega$ is the optical frequency, $\hbar$ is the reduced Planck's constant, 
$\alpha,\beta = x,y$ are Cartesian tensor labels, $n,m$ are band indices, $\ket{ n \bk}$ is a Bloch state, 
$E_{n\bk}$ is a band energy, $f_{n\bk}$ is Fermi-Dirac band occupation probability, and $\eta$ is a disorder 
broadening parameter that is set to 0.5 meV.  
In the Dirac cone model \cite{Lei2020} $\partial_{x} H_{\bk} = \hbar v_D \sigma_y \tau_z$, $\partial_{y} H_{\bk} = - \hbar v_D \sigma_x \tau_z$ where $\sigma_{\alpha}$ is a Pauli matrix acting on spin. 
The band energies in Eq.~\ref{eq:optical_conductivity} depend only on the magnitude of wavevector $\bk$, and the 
velocity matrix elements have a simple angle dependence that allows angular integrals to be performed analytically.  
Our calculations are therefore performed by integrating numerically over the radial direction of 2D k-space after performing the angular integrals analytically. 
Because we employ continuum Dirac models for the $\hat{x}-\hat{y}$ planes, the numerical integrals require 
a k-space cutoff. In our numerical calculations we used an adaptive $k$-mesh with a higher density
over the range of $k$ with a large Berry curvature, and a sparse mesh where the Berry curvature is small. 
In this specific model, the Berry curvatures are mainly contributed around $\bk = 0$ 
and our k-mesh samplings are set as follows: 
\begin{equation}
    k \in \begin{cases}
       [0,2^{n}] \frac{\pi}{a} & n = 0\\
       [2^{n-1},2^{n}] \frac{\pi}{a}& n = 1,2,3,..., M\\
    \end{cases}, 
\end{equation}
Here $a$ is selected so that the Berry curvature is concentrated in the range of $[0,\pi/a]$
and therefore depends on the band gap where band inversion appears. 
$M$ determines the cutoff of the wavevector.  We normally set $M=12$ which places any 
anomalies associated with the cut-offs outside of the range of frequencies that we plot.  


\ifpdf
\begin{figure}[htp!]
  \centering
  \includegraphics[width=0.9 \linewidth ]{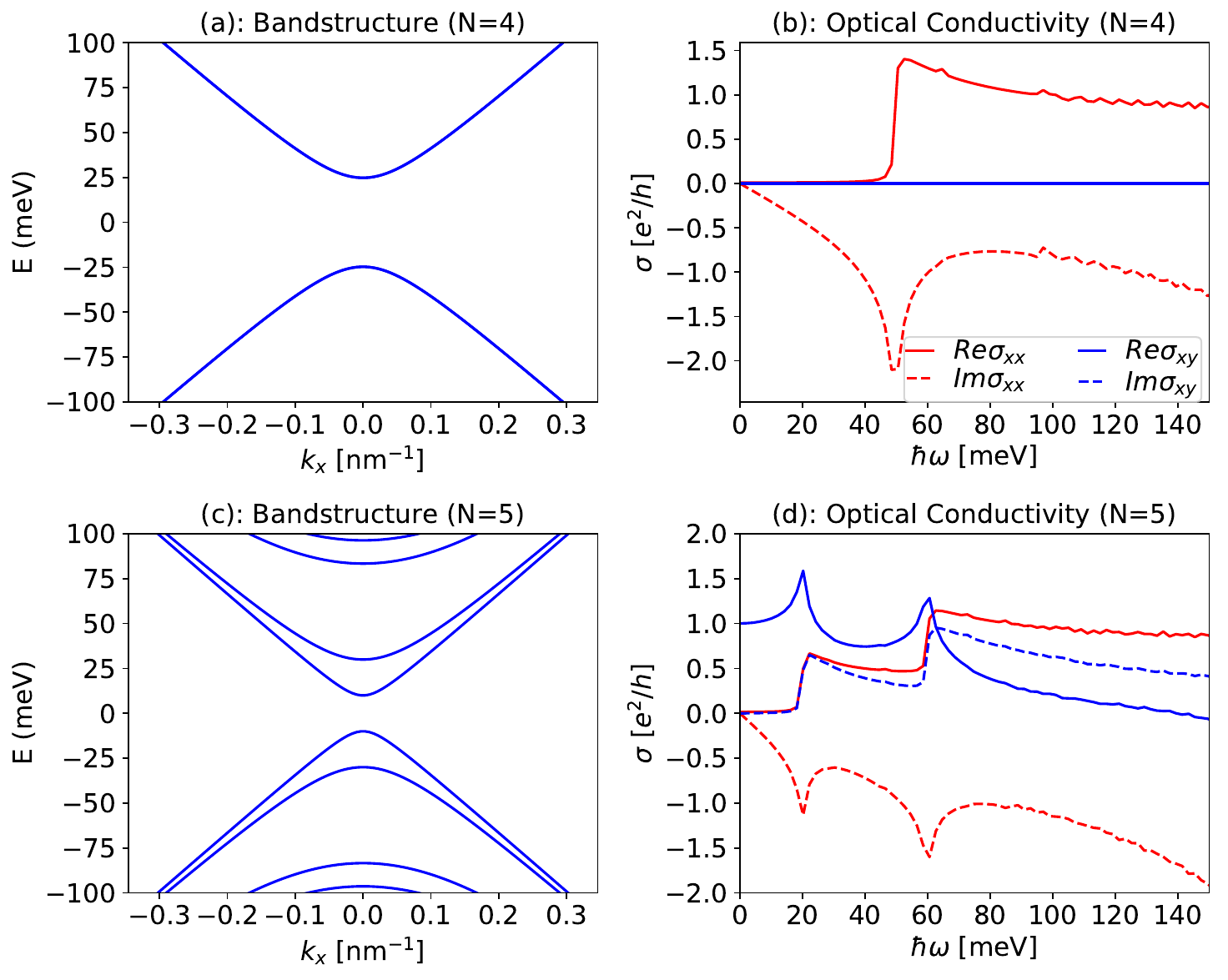}
  \caption{Bandstructure and optical conductivity for antiferromagnetic MnBi$_2$Te$_4$ thin films with  
  septuple layer numbers $N=4,5$.
  (a) Bandstructure of $N=4$ MnBi$_2$Te$_4$ along x-direction of the momentum, with $k_y$ to be zero since the Hamiltonian is rotational invariant.  Even $N$ bands are doubly degenerate due to the $\mathcal{T} \mathcal{I}$ symmetry discussed in the text.
  (b) Two-dimensional optical conductivity {\it vs.} frequency of $N=4$ MnBi$_2$Te$_4$.  
  The Hall conductivity vanishes identically due to $\mathcal{T} \mathcal{I}$ symmetry.   
   (c) and (d) Bandstructure and conductivity plots for $N=5$ MnBi$_2$Te$_4$, which has a 
   quantum anomalous Hall effect.  The {\it dc} limit of the 
   Hall conductivity is therefore quantized.
   Longitudinal conductivities ($\sigma_{xx}$) are plotted in red and 
   Hall conductivities ($\sigma_{xy}$) in blue.  In each case, the real and imaginary parts are 
   plotted as solid and dashed lines.  All conductivities are expressed in units of $e^2/h$.
   }
  \label{fig:band_conductivity}
\end{figure}
\fi

In Fig. \ref{fig:band_conductivity} we illustrate the main features of 
$\mathcal{E}_z=0$ systems by plotting 
the bandstructure and the frequency-dependent longitudinal and Hall 
optical conductivities  ($\sigma_{xx}(\omega)$ and $\sigma_{xy}(\omega)$) of antiferromagnetic MnBi$_2$Te$_4$ thin films with septuple-layer numbers $N=4$ and $N=5$ (The optical conductivities for other antiferromagnetic thin films as shown in Fig. \ref{fig:conductivity_AF} of appendix \ref{appendix:optical_sigma}).  
The ground states for $N=4$ and other even $N$ films are referred to as axion insulators in the literature, and have similar conductivities.  
The $N=5$ case is an example of an odd $N$ magnetic configuration that supports a quantum anomalous Hall effect 
and therefore does not have axion electrodynamics.   
The electronic structure of $N=4$ thin film is invariant under combined time-reversal ($\mathcal{T}$) and inversion ($\mathcal{I}$) symmetry, 
which leads, via a generalized Kramer's theorem, to the doubly degenerate 2D bands \cite{Lei2021_metamagnetism} shown in Fig. \ref{fig:band_conductivity} (a).
In this case, as shown in Fig. \ref{fig:band_conductivity} (b), both real (red solid curve) and imaginary (dashed red curve) 
parts of longitudinal conductivity ($\sigma_{xx}$) approach 0 in the low frequency limit and have 
interband features at $\approx 50$ meV. 
The Hall conductivity ($\sigma_{xy}$), and therefore both Kerr and Faraday angles,
vanish identically over the entire range of frequencies due to $\mathcal{T}\mathcal{I}$ symmetry.
For thin films with an odd number of layers, there is no $\mathcal{T} \mathcal{I}$ symmetry 
and the band degeneracy is lifted as shown in Fig. \ref{fig:band_conductivity} (c)). 
Antiferromagnetic MnBi$_2$Te$_4$ thin films with thickness $N>3$ are 
Chern insulators
with Chern number $C = 1$ and therefore have quantized {\em dc} Hall conductivities
as shown in Fig. \ref{fig:band_conductivity} (d).
The gap of the $N=5$ Chern insulator is small ($\sim 20$ meV) because the minimum thickness 
necessary for Hall quantization is only modestly exceeded \cite{Lei2021_QAH}.
In Fig. \ref{fig:conductivity_FM} of the appendix \ref{appendix:optical_sigma} we summarize the properties of the conductivity tensor in thin films with 
spin-aligned magnetic configurations, which can be induced in 
MnBi$_2$Te$_4$ by applying magnetic fields exceeding $\sim 5$ Tesla \cite{Otrokov2019,Deng_2020,Ge2020}. 
Based on these optical conductivities, the corresponding Kerr and Faraday rotations {\it vs.} optical frequencies of MnBi$_2$Te$_4$ thin films are estimated as shown in Fig. \ref{fig:Kerr_Faraday}, 
here we estimate the magneto-optical rotation angles in the 2D limit. 
The Faraday and Kerr rotations for AF thin films are shown in Fig. \ref{fig:Kerr_Faraday} (a) and (c), from which we see that for thin films with even-number-layers (corresponds to dashed curves), there is no Faraday and Kerr signals as $\sigma_{xy} (\omega) = 0$. 
For odd N thin films, however, there is a large Faraday and Kerr rotation angle ($\theta_{F/K}$) at finite frequencies even though the MnBi$_2$Te$_4$ thin films are trivial insulators, see $N =1$ or $N = 3$ thin films for example, although $\theta_{F/K}$ is still 0 in the DC limit. When the thin films are in Chern insulator states, the Faraday and Kerr rotation angles in the DC limit approach to the quantized value, i.e. $-tan^{-1} (1/4\pi \Re\sigma_{xy}) \approx - \pi/2$ for Kerr rotation angle; and $tan^{-1} (4\pi \Re\sigma_{xy}) = C tan^{-1} \alpha $ for Faraday rotation angle. Here $C$ is the Chern number and $\alpha$ is the fine structure constant. 

\ifpdf
\begin{figure}[htp!]
 \centering
 \includegraphics[width=0.9 \linewidth ]{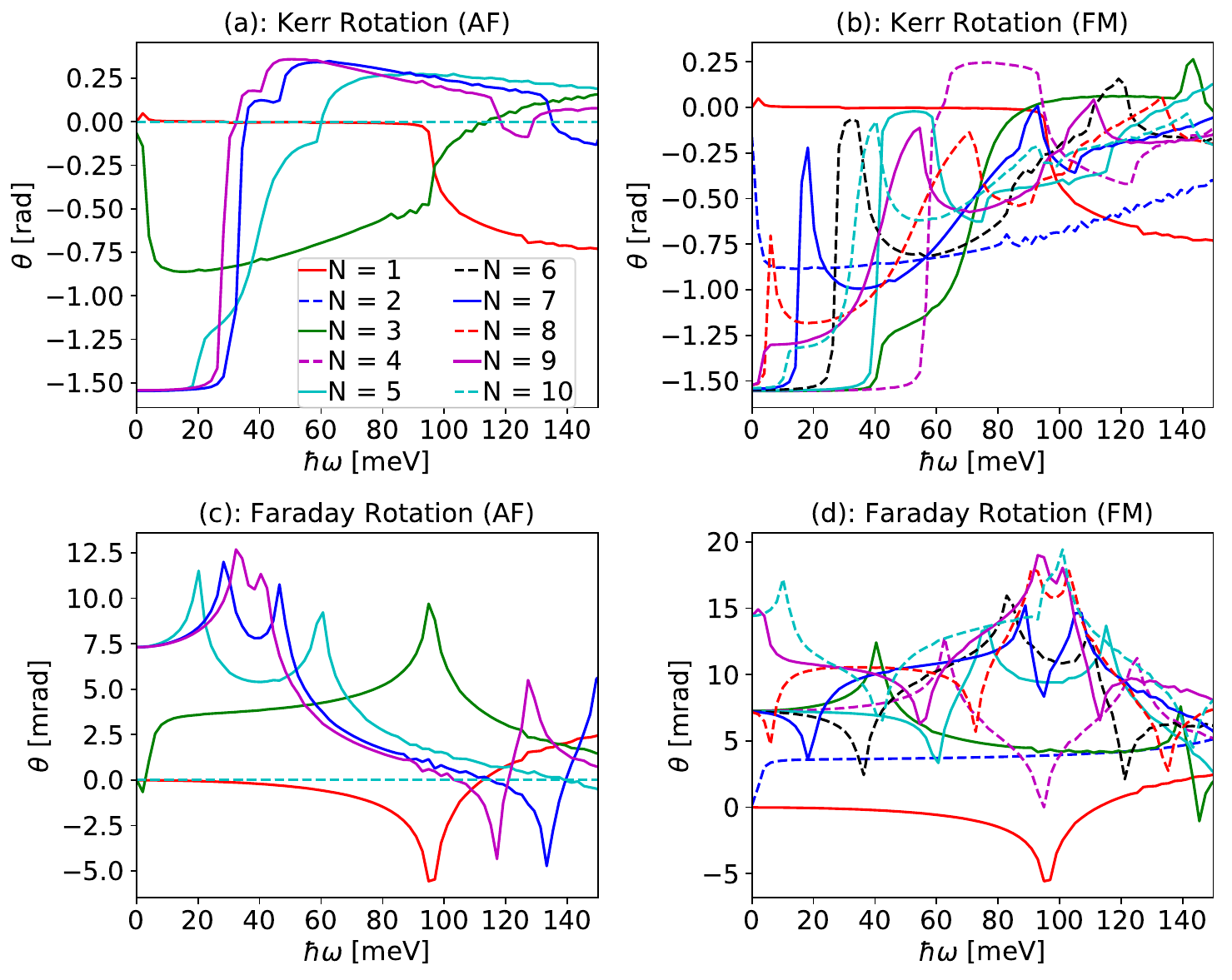}
 \caption{Kerr and Faraday rotation angle {\it vs.} optical frequency.
 (a) and (b) are the plots of Faraday rotation angles of MnBi$_2$Te$_4$ thin films in AF and FM states;
  (b) Faraday rotation angles of MnBi$_2$Te$_4$ thin films in FM state;
 While (c) and (d) are the plots of Kerr rotation angles.
  of MnBi$_2$Te$_4$ thin films in AF state;
  (d) Kerr rotation angles of MnBi$_2$Te$_4$ thin films in FM state;
 In these plots, the same color labels the same thickness of MnBi$_2$Te$_4$ thin films in all the 4 panels. 
 }\label{fig:Kerr_Faraday}
\end{figure}
\fi

\section{Electric-field dependence of the Faraday and Kerr angles}
The $\mathcal{TI}$ symmetry that causes the Kerr, Faraday
and orbital magnetization responses to simultaneously vanish in axion insulator states
is broken by an electric field $\mathcal{E}_z$ applied
across the film.  
Based on Schr\"odinger-Poisson equation, we calculate the orbital magneto-electric response for MBT thin films with the modern theory of orbital magnetization, which leads to the following expression
\cite{Xiao2005,Ceresoli2006,Thonhauser2005,Shi2007}:
\begin{equation}
\begin{split}
    M_{orb} = \frac{e}{2\hbar} \int & \frac{1}{(2\pi)^2} d\bk 
              \sum_{n} f_{n\bk} \rm{Im} \langle \partial_{\bk} u_{n\bk} |\\
              & \times(H_{\bk} + E_{n\bk} - 2\mu)| \partial_{\bk} u_{n\bk} \rangle ,
\end{split}
\end{equation}
where $f_{n\bk}$ is the Fermi-Dirac distribution function, $H_\bk$ is the Hamiltonian introduced in Eq. \ref{eq:model}, $E_{n\bk}$ is the eigenvalue of the $n^{th}$ subbands, $\mu$ is the chemical potential, and the wavevector integrals are 
over two-dimensional momentum space. 
The orbital magnetism response to $\mathcal{E}_z$ (denoted as $\mathcal{E}$ in the plots since we consider the electric field only in z-direction), plotted as blue curves in Fig. \ref{fig:kerr_field_om})(a) and (b),
is initially linear with small finite thickness corrections \cite{pournaghavi2021non} to the 
quantized response coefficient, which is common to all topological insulators.
This topological response is strong, at least compared to that of typical magneto-electric materials like Cr$2$O$_3$.  

\ifpdf
\begin{figure}[htp!]
  \centering
  \includegraphics[width=0.95 \linewidth ]{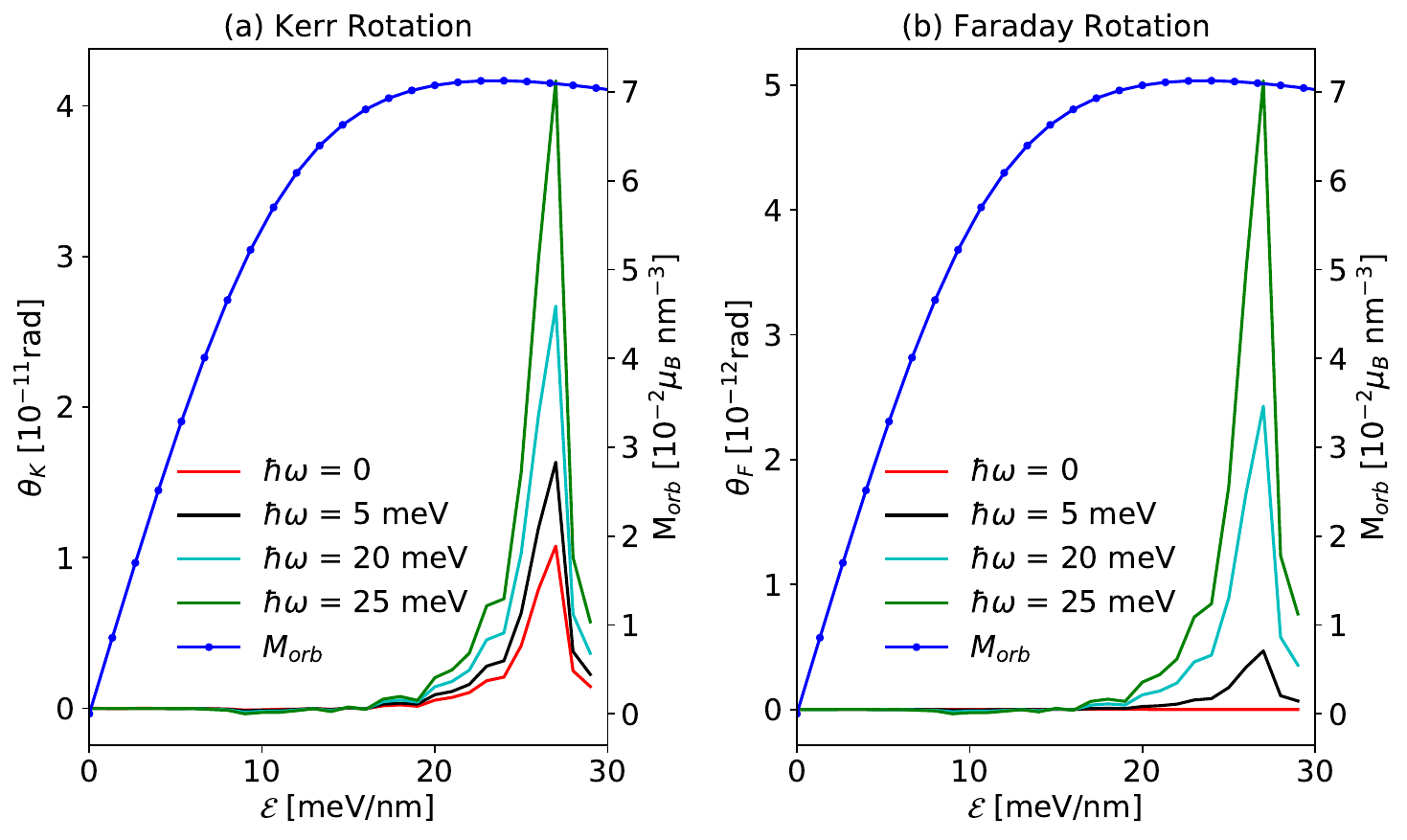}
  \caption{Comparison of the electric field($\mathcal{E}$, corresponding to $\mathcal{E}_z$ in the main text) dependence of the orbital magnetization, and the Kerr (a) and Faraday (b) rotation angles
  for antiferromagnetic MnBi$_2$Te$_4$ thin films with a thickness of $N=4$ septuple layers.  In this figure 
  the Kerr and Faraday angles are calculated in 2D limit in which the film thickness is very small compared to the 
  light wavelength.
  }\label{fig:kerr_field_om}
\end{figure}
\fi

We now combine the simplified coupled Dirac cone model and self-consistent Schr\"odinger-Possion equations\cite{Lei2021_QAH} 
to model the effect of $\mathcal{E}_z$ on the magneto-optical response. 
In Fig. \ref{fig:kerr_field_om} (a) and (b) we show typical 2D limit Kerr and Faraday rotation angles for 
even-layer thin films calculated from the conductivity tensor
of an $N = 4$ axion insulator thin film using a substrate dielectric constant $n_s=1$.
We see that the Kerr and Faraday angles have extremely
small linear response coefficients, and that 
they remain small even when the vertical electric field $\mathcal{E}_z$ \cite{Lei2021_QAH} 
is near the critical value at which the film gap vanishes.
We attribute these very small values to the approximate locality of the Hall response at 
top and bottom surfaces.  Because the surface magnetizations are opposite the total Hall conductivities 
nearly vanish at all frequencies even for $\mathcal{E}_z \ne 0$.
The Kerr and Faraday angles in Fig.~\ref{fig:kerr_field_om} ($\sim 10^{-11}$ rad) 
lie below current Kerr angle detection limits\cite{Xia2006_Kerr,Rowe2017_Faraday}, to the best of our knowledge.
The frequency and electric field dependence of the underlying conductivities is 
presented in Fig. \ref{fig:conductivity_xx_field_frequency}-\ref{fig:kerr_field} of the 
appendix \ref{appendix:sigma_efield}.  Because these response coefficients 
calculated in the 2D limit are practically zero even at $\mathcal{E}_z \ne 0$, the finite 
thickness corrections we examine next are actually dominant. 

\section{Thickness Dependence}
Because the Kerr response is so weak in the 2D limit, finite thickness corrections can easily be important.  
To assess the film thickness dependence of the Kerr and Faraday response, we first model the thin
film as two Dirac surfaces that support half-quantized Hall effects of opposite signs and are 
separated by a dielectric bulk. The electromagnetic wave then scatters at both interfaces.
Denote the incoming field as $[\Tilde{E^{ti}} ~~ \Tilde{E^{rj}}]^T $ and outgoing field as $[\Tilde{E^{ri}} ~~ \Tilde{E^{tj}}]^T $, 
at the interface  the incoming and outgoing fields are connected with the scattering matrix $S$ which reads as:
\begin{equation}
    S = 
    \begin{pmatrix}
\Bar{r} & \Bar{t'}\\
\Bar{t} & \Bar{r'} \\
\end{pmatrix},
\end{equation}
Here $\bar r'$ and $\bar t'$, the reflection and transmission tensors for incidence from the right,
are obtained from $\bar r$ and ($\bar t$) by reversing the wavevector direction 
and interchanging the dielectric constants on opposite sides of the interface \cite{Tse2011}.
$\Bar{r}$, $\Bar{t}$ are defined as:
\begin{equation*}
    \Bar{r} = \begin{pmatrix}
        r_{xx} & r_{xy} \\
        -r_{xy} & r_{yy}
    \end{pmatrix}, ~~
    \Bar{t} = \begin{pmatrix}
        t_{xx} & t_{xy} \\
        -t_{xy} & t_{yy}
    \end{pmatrix}.
\end{equation*}
This total reflection and transmission tensors $\bar r$ and $\bar t$ 
can be composed from the top ($T$) and bottom ($B$) single-interface scattering matrices: 
\begin{equation}\label{reflection_transmision_tensor}
\begin{aligned}
    & \bar r = \bar r_T + \bar t'_T \bar r_B (1-\bar r'_T \bar r_B)^{-1} \bar t_T, \\
    & \bar t = \bar t_B (1- \bar r'_T \bar r_B)^{-1} \bar t_T.
\end{aligned}
\end{equation}
The resulting Kerr and Faraday rotations depend strongly on the dielectric constants 
experienced by incoming and outgoing light. 
If we assume that light is incoming from vacuum with relative 
dielectric constant ($n_0=1$) and is 
outgoing to an infinite substrate with relative dielectric constant $n_s$, 
the Faraday rotation angle for an antiferromagnetic thin film is non-zero only when $n_s \ne 1$.
In contrast, the Kerr rotation angle is non-zero when $n_s = n_0$, as illustrated in Fig. \ref{fig:kerr_vs_dielectric} (a) and (b). 
In Fig. \ref{fig:kerr_vs_dielectric} (c) and (d) the Kerr and Faraday rotation angles for thin films with FM state are plotted.

\ifpdf
\begin{figure}[htp!]
  \centering
  \includegraphics[width=0.9 \linewidth ]{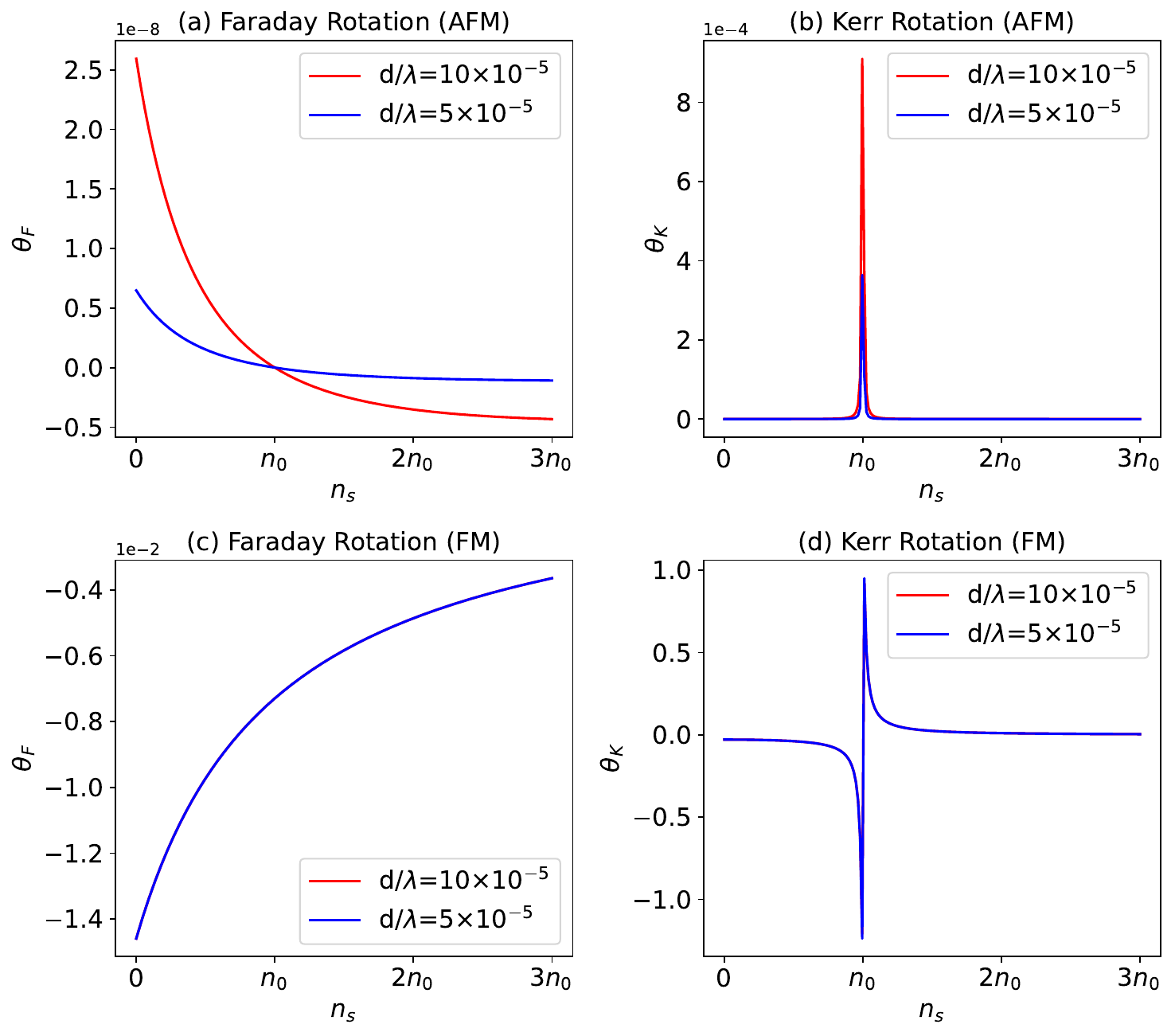}
  \caption{ Dependence of magneto-optical Faraday and Kerr rotation angles {\it vs.} the relative dielectric constant of the substrate.
  (a) and (b) are the dependence of Faraday/Kerr rotation angles on relative dielectric constant of substrate for AF states,
  while (c) and (d) are the plots for FM state.
  }\label{fig:kerr_vs_dielectric}
\end{figure}
\fi



When the thickness of substrates is considered to be finite, i.e. with substrate thickness $d_s$,
We can account for both the thickness and the index of refraction ($n_s$) of the 
substrate by replacing $\bar t_B$ and $\bar r_B$ by $\bar t_S$ and $\bar r_S$
which can be calculated by composing two single-interface scattering matrices - 
in this case the interfaces between substrate and sample and between substrate and vacuum:
\begin{equation}
\begin{aligned}
    & \bar r = \bar r_T + \bar t'_T \bar r_{SB} (1-\bar r'_T \bar r_{SB})^{-1} \bar t_T, \\
    & \bar t = \bar t_{SB} (1- \bar r'_T \bar r_{SB})^{-1} \bar t_T,
\end{aligned}
\end{equation}
where $r_{SB}$ and $t_{SB}$ are calculated as:
\begin{equation}
\begin{aligned}
    & \bar{r}_{SB} = \bar r_B + \bar t'_B \bar r_S (1-\bar r'_B \bar r_S)^{-1} \bar t_B, \\
    & \bar{t}_{SB} = \bar t_S (1- \bar r'_B \bar r_S)^{-1} \bar t_B.
\end{aligned}
\end{equation}
Here $r_B$/$t_B$ is the scattering matrix in the interfaces between substrate and sample and $r_S$/$t_S$ is the one between substrate and vacuum.

The Kerr signal can thus also depend on the substrate thickness $d_s$ 
when accounting for both the thickness and the index of refraction ($n_s$) of the 
substrate, as illustrated in Fig. \ref{fig:Kerr_Faraday_Thickness} (a).
The dependence of the Kerr and Faraday rotation angles on $d_s$ and $n_s$ 
are illustrated in Fig. \ref{fig:Kerr_Faraday_Thickness} (b),
for the case of an optical frequency close to the gap (25 meV) and an electric field of 25 meV/nm.
This $\mathcal{E}_z$ is smaller than the critical electric field, 
The substrate index of refractions in Fig. \ref{fig:Kerr_Faraday_Thickness} (b) are 
$n_s = 3$ a typical value for hexagonal boron nitride, 
$n_s = 5$ a typical value for silicon dioxide,
and $n_s = 10$ a typical value of aluminum oxide.  These results demonstrate explicitly that 
reflection off the substrate-vacuum interface is important whenever the light is not absorbed 
in the substrate. 
With the substrate thickness $d_s=200 \lambda$ as an example, the dependence of Kerr and Faraday rotations
on the thickness of sample is plotted in Fig. \ref{fig:Kerr_Faraday_Thickness} (c) and (d), which is linearly 
proportional with the thickness of sample, and the slopes depend on the substrate index of refractions.

\ifpdf
\begin{figure}[htbp!]
  \centering
  \includegraphics[width=1.0 \linewidth ]{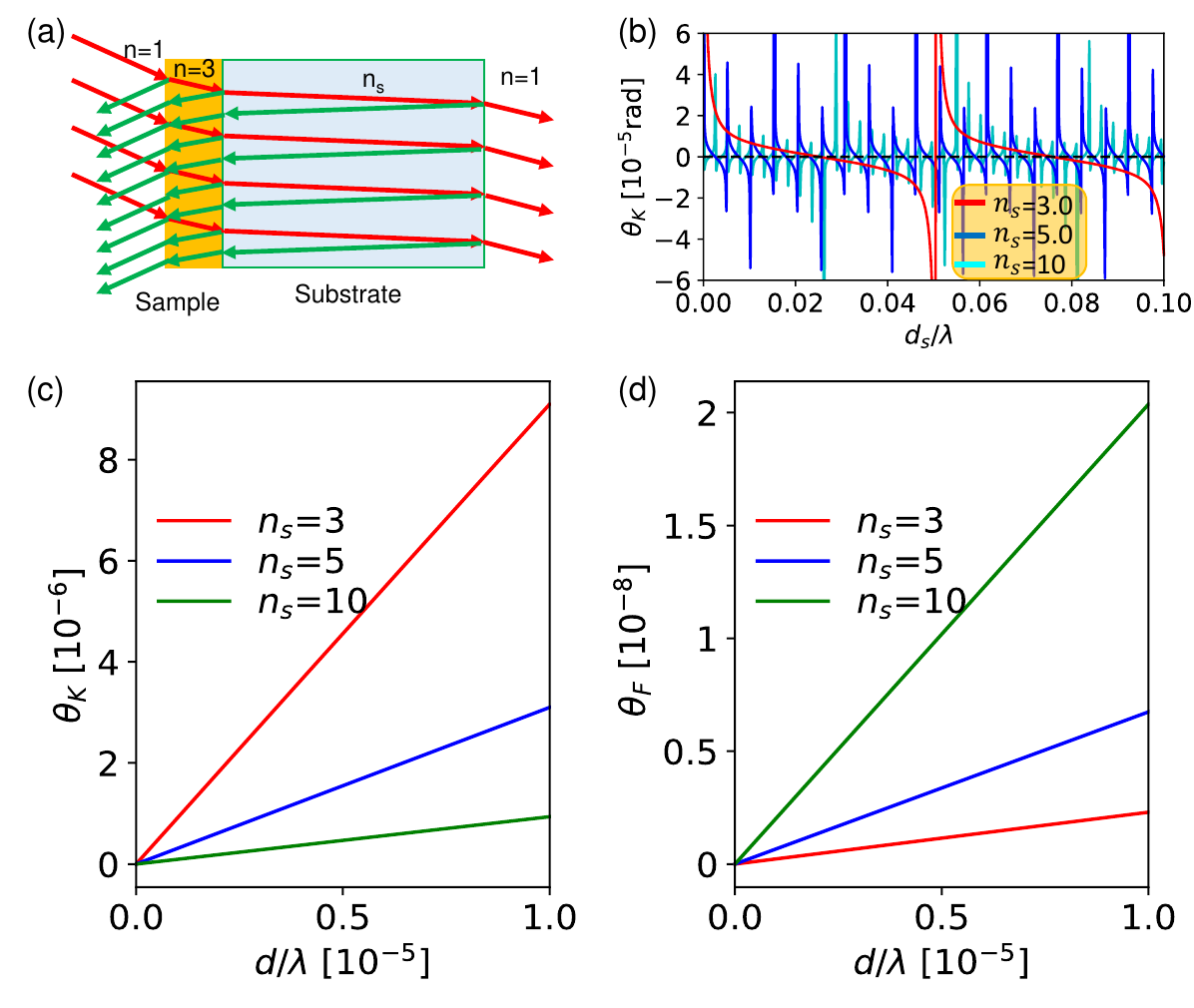}
  \caption{
  (a) Illustration of light reflection and transmission for a finite-thickness film with a substrate;
  (b) Kerr rotation {\it vs.} substrate thickness $d_s$ for various substrate dielectric constants.
  These results were calculated for a light frequency of 25 meV, which is below the gap at $\mathcal{E}_z=0$, 
  an electric field $\mathcal{E}_z =25$ meV/nm, which is close to but smaller 
  than the critical electric field for gap closure, and a $N=4$ septuple layer 
  sample thickness.  (c)-(d): Dependence of Kerr and Faraday angles on
  sample thickness at $\mathcal{E}_z=0$ with the substrate thickness set to $d=200\lambda$.
  The range of thickness to wavelength ratio corresponds to photon energies $\sim 10$ meV and 
  films with fewer than $\sim 10$ septuple layers.
  }
  \label{fig:Kerr_Faraday_Thickness}
\end{figure}
\fi

Finally we examine the role of 
imperfect compensation between the half-quantized hall conductivities on the 
top and bottom surfaces, by fixing the difference of Hall conductivity between the two surfaces at
$\sigma_T - \sigma_B = e^2/h$ and allowing the total to be non-zero.  Our explicit calculations 
discussed previously have shown that 
the total Hall conductivity $\sigma_{xy}^{T}$ remains extremely small (as illustrated in Fig. \ref{fig:conductivity_xy_field_frequency}) even for 
$\mathcal{E}_z \ne 0$ as long as the 2D system is an insulator.  We do expect sizable 
conductivities to arise, however, when the Fermi level is in the local gap on 
one side of the film and not on the other.   
In Fig. \ref{fig:Kerr_Faraday_Sigma}(a) and (b) we plot the Kerr and 
Faraday angles {\it vs.} $\sigma_{xy}^T$ for substrate thickness $d_s=200 \lambda$
for both $n_s=3$ and $n_s=10$.  In each case we plot results obtained 
using the 2D approximation and for the thicknesses of $N=4$ and $8$ septuple layer films.
These calculations demonstrate that the Kerr and Faraday angles have additive linear 
contributions from both finite thickness and from a finite value of the total 
Hall conductivity of the film, and that both angles are sensitive to substrate properties.

\ifpdf
\begin{figure}[htbp!]
  \centering
  \includegraphics[width=1.0 \linewidth ]{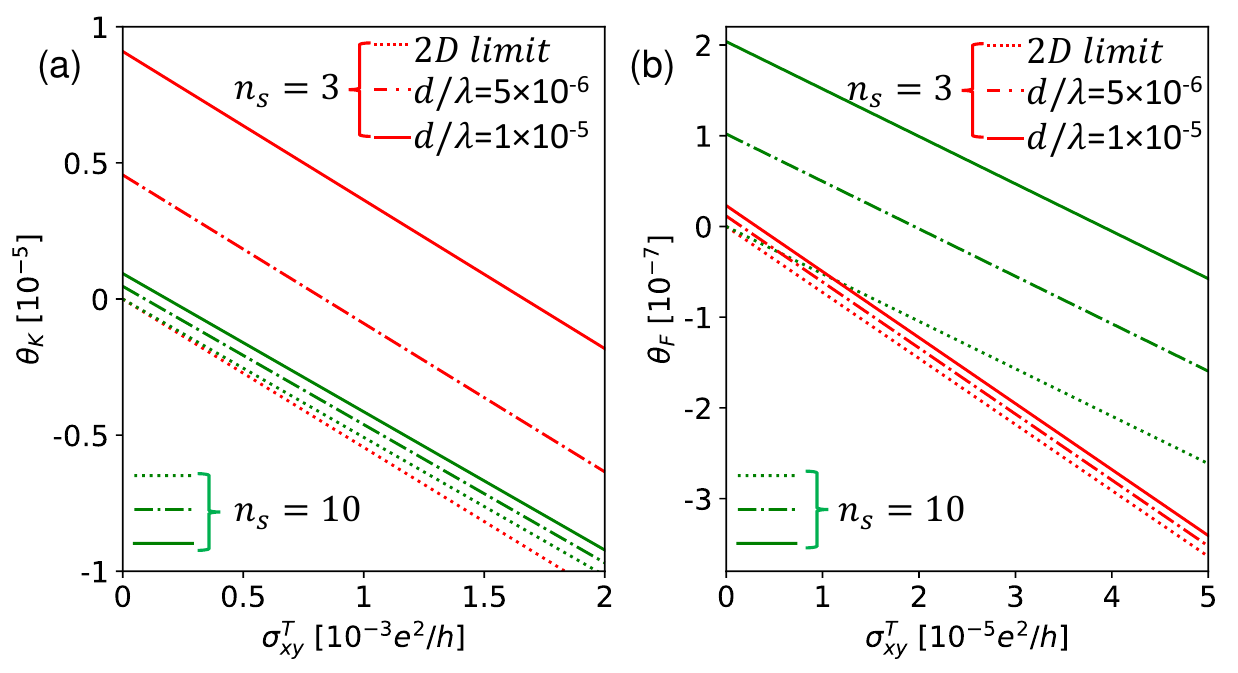}
  \caption{Dependence of Kerr and Faraday angles on total Hall conductivity 
  with substrate thickness set to $d_s = 0.56 \lambda$.  These results demonstrate
  additive linear dependence on $\sigma_{xy}^T$ and sample thickness $d$ that are distinctly
  different for substrate index of refraction $n_s=3$ (red) and $n_s=10$ (green).  
  }
  \label{fig:Kerr_Faraday_Sigma}
\end{figure}
\fi

\section{Discussion} 
MnBi$_2$Te$_4$ thin films with an odd number $N$ of septuple layers are two-dimensional insulating ferromagnets
that exhibit the quantum anomalous Hall effect.  Because they are ferromagnets with strong spin-orbit coupling, they 
have non-zero spin and orbital magnetizations in the absence of any external fields.  Because they exhibit the quantum anomalous Hall effect, the films have large total optical Hall conductivities at frequencies below the gap that lead to 
substantial Kerr and Faraday effects that are readily observable, and are indirectly related to the topological 
magneto-electric effect.  In this paper we focus on the magneto-electric and magneto-optical response
properties of even $N$ thin films, which are axion insulators.  
We find that although the {\em dc} magnetization response to electric field is quantized, the response of the magneto-optical Faraday and 
Kerr and angles to an electric field is extremely weak, and what survives might be difficult to disentangle.
We predict that at frequencies below the gap, the Kerr and Faraday angles of realistic thin films
will have additive small contributions from the finite thicknesses of the MnBi$_2$Te$_4$ samples, and from 
imperfect compensation between contributions to the total Hall conductivity from the top and bottom of the thin films.
The best way to identify the topological magneto-electric effect, we believe, is to do it thermodynamically by
measuring the temperature and magnetic-field dependent capacitance of hBN-encapsulated MnBi$_2$Te$_4$ thin films.

\section{Acknowledgement}
This work was sponsored by the Department of Energy under grant DE-SC0021984.  

\appendix
\setcounter{figure}{0}
\renewcommand{\figurename}{Fig.}
\renewcommand{\thefigure}{A\arabic{figure}}

\section{Optical conductivities}\label{appendix:optical_sigma}

The optical conductivities of antiferromagnetic (AF) MnBi$_2$Te$_4$ thin films from $N=1$ septuple layer to $N=10$ septuple layers are 
illustrated in Fig. \ref{fig:conductivity_AF}.  The real and imaginary parts of $\sigma_{xx}(\omega)$ and $\sigma_{xy}(\omega)$ 
are plotted separately in panels (a)-(d). 
In Fig. \ref{fig:conductivity_AF} (a) and (b) we see that 
both the real and imaginary parts of $\sigma_{xx}(\omega)$ have typical interband absorption edge features
at the $N$-dependent gap energy. The real and imaginary parts of $\sigma_{xy}(\omega)$ vanish identically for all even $N$ films, as shown in Fig. \ref{fig:conductivity_AF} (a) and (b), due to the combined time-reversal times spacial inversion symmetry. 
The quantum anomalous Hall (QAH) and trivial insulators are clearly distinguished by the Hall conductivities at 
frequencies well below the band gap.

\ifpdf
\begin{figure}[htp!]
  \centering
  \includegraphics[width=0.9 \linewidth ]{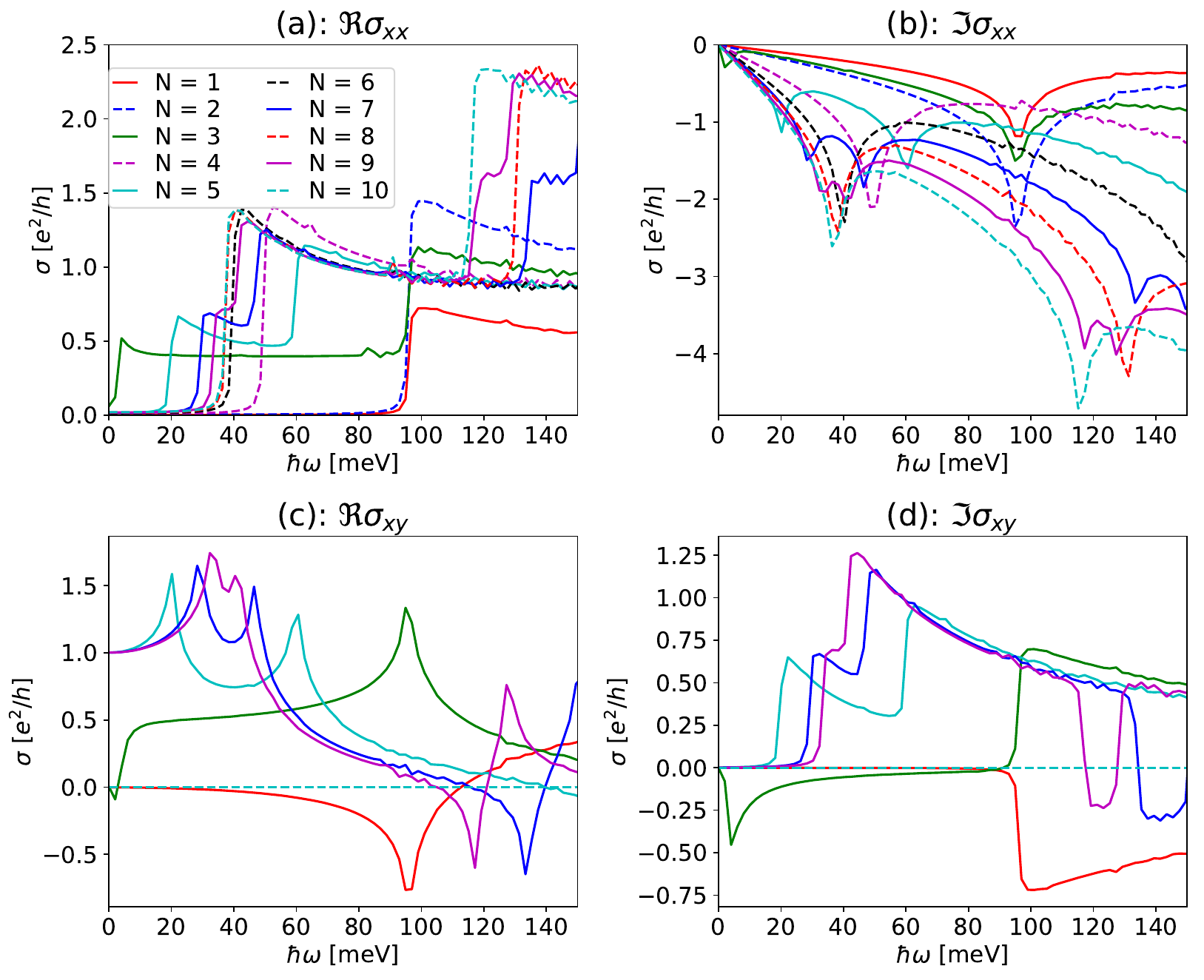}
  \caption{Optical conductivity tensors of antiferromagnetic thin film with septuple layer numbers $N$ varying from 1 to 10.  
  (a) Real parts of $\sigma_{xx} (\omega)$;
  (b) Imaginary parts of $\sigma_{xx} (\omega)$; 
  (c) Real parts of $\sigma_{xy} (\omega)$;
  (d) Imaginary parts of $\sigma_{xy} (\omega)$.
  The same colors are used to label the MnBi$_2$Te$_4$ thickness in panels (b)-(d) as in 
  panel (a).
  }\label{fig:conductivity_AF}
\end{figure}
\fi

For MnBi$_2$Te$_4$ thin films in ferromagnetic (FM) state, the frequency-dependence of longitudinal optical conductivity $\sigma_{xx}(\omega)$ is similar to the ones in AF state. In Fig. \ref{fig:conductivity_FM} (a) and (b) we show the plots of real and imaginary parts of $\sigma_{xx}(\omega)$ for MnBi$_2$Te$_4$ thin films with thickness from 5 SLs to 10 SLs, the step-like behaviors of $\Re \sigma_{xx}(\omega)$ origin from the accumulated exited subbands when the frequency increases. The Hall conductivities, shown in Fig. \ref{fig:conductivity_FM} (c) and (d) where we use the same color to label the same thickness as in panels (a) and (b), have different dependence on optical conductivity compared with the AF state. In the DC limit, a topological phase transition happens when the thickness increases to $N \ge 9$, above the critical thickness the Chern number increases from 1 to 2. 
$\Im \sigma_{xy}(\omega)$ shown in Fig. \ref{fig:conductivity_FM} (d) indicates that the frequency dependence differs between $N \ge 9$ and $N < 9$ thin films. Note that $\Im \sigma_{xy}(\omega) < 0$ at low frequencies for $N < 9$ thin films, this is due to the negative Berry curvature close to $\Gamma$ point in 2D bands. The negative Berry curvature also leads to a decrease of $\Re \sigma_{xy}(\omega)$ shown in Fig. \ref{fig:conductivity_FM} (c).

\ifpdf
\begin{figure}[htbp!]
  \centering
  \includegraphics[width=0.9 \linewidth ]{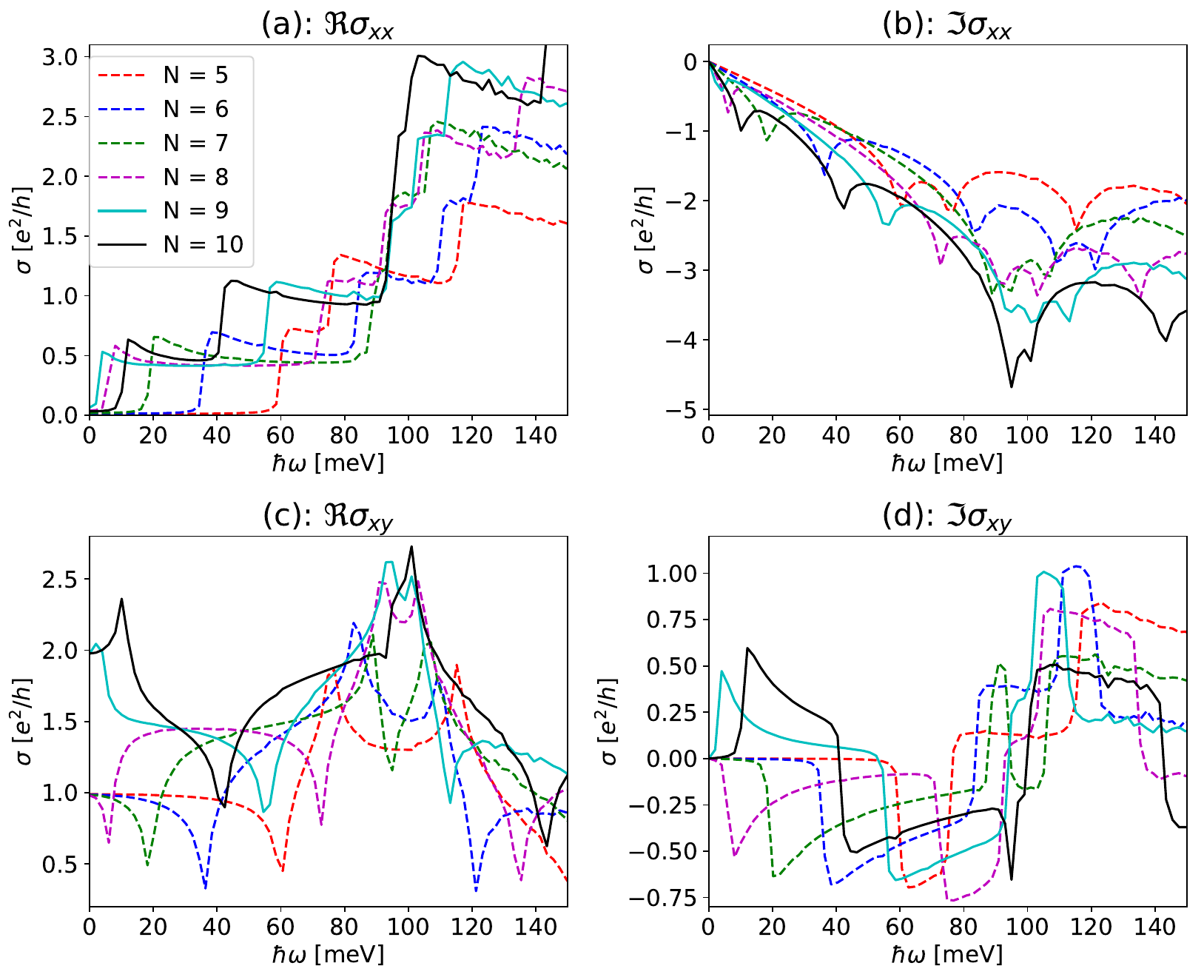}
  \caption{Optical conductivities for ferromagnetic thin films with the number of layers from 5 to 10.
  (a) Real parts of $\sigma_{xx} (\omega)$;
  (b) Imaginary parts of $\sigma_{xx} (\omega)$; 
  (c) Real parts of $\sigma_{xy} (\omega)$;
  (d) Imaginary parts of $\sigma_{xy} (\omega)$.
  In panel (c)-(d), the same colors as panel (a) are used to label the thickness of MnBi$_2$Te$_4$ thin films.  
  }\label{fig:conductivity_FM}
\end{figure}
\fi

\section{Electric field dependence of optical conductivity}\label{appendix:sigma_efield}

In Fig. \ref{fig:conductivity_xx_field_frequency} the dependence of optical conductivities $\sigma_{xx}$ on electric field and frequency 
for 4-layer antiferromagnetic thin film are plotted.

\ifpdf
\begin{figure}[htbp!]
  \centering
  \includegraphics[width=0.9 \linewidth ]{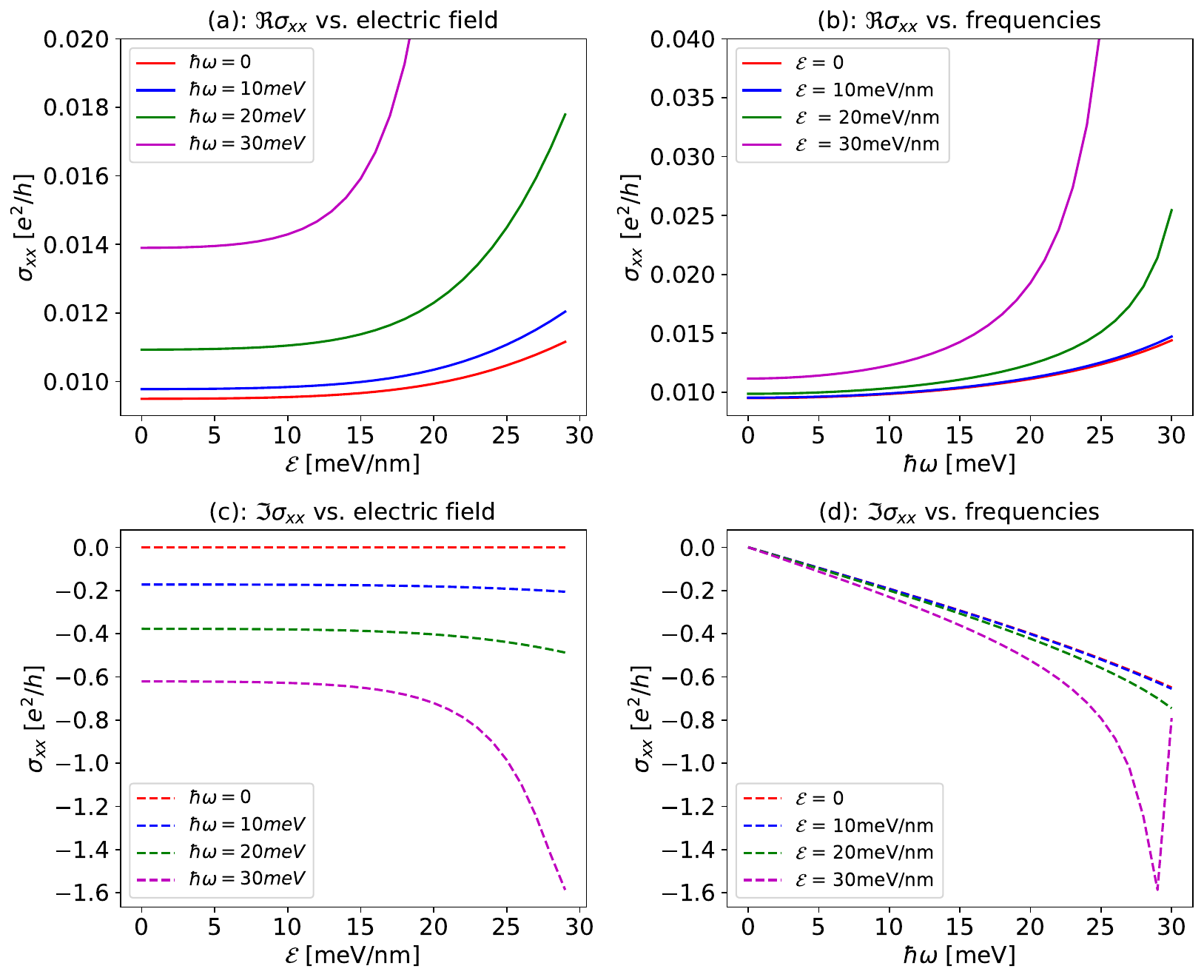}
  \caption{Optical conductivities $\sigma_{xx}$ for 4-layer antiferromagnetic thin film.
  (a) Real parts of $\sigma_{xx}$ {\it vs.} electric field;
  (b) Real parts of $\sigma_{xx}$ {\it vs.} optical frequencies $\omega$; 
  (c) Imaginary parts of $\sigma_{xx}$ {\it vs.} electric field;
  (d) Imaginary parts of $\sigma_{xx}$ {\it vs.} optical frequencies $\omega$; 
  }\label{fig:conductivity_xx_field_frequency}
\end{figure}
\fi

The dependence of optical conductivities $\sigma_{xy}$ electric field and frequency are shown in Fig. \ref{fig:conductivity_xy_field_frequency}.
\ifpdf
\begin{figure}[htbp!]
  \centering
  \includegraphics[width=0.9 \linewidth ]{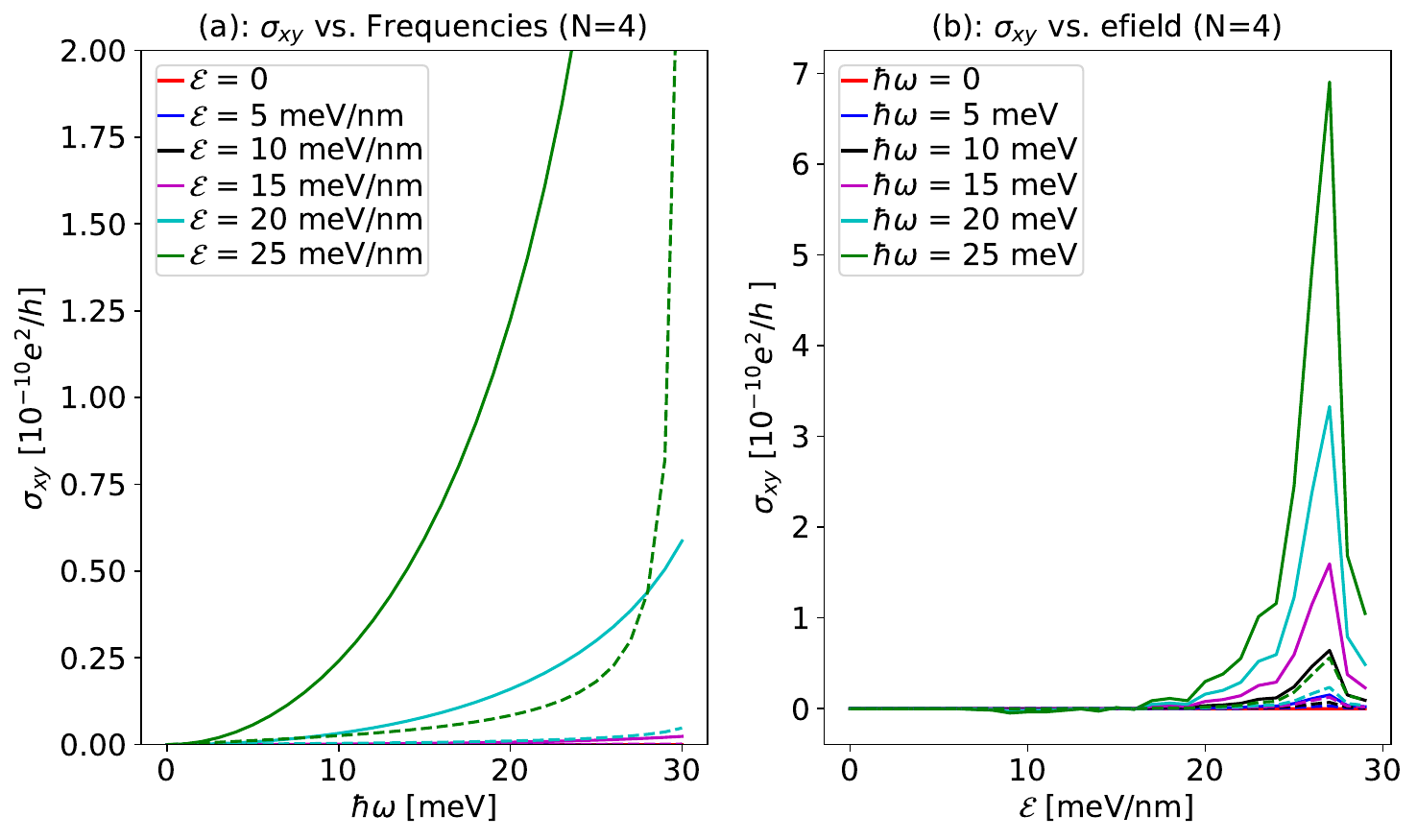}
  \caption{Optical Hall conductivities $\sigma_{xy}$ for 4-layer antiferromagnetic thin film.
  (a) Real and imaginary parts of $\sigma_{xy}$ {\it vs.} optical frequencies $\omega$;
  (b) Real and imaginary parts of $\sigma_{xy}$ {\it vs.} electric field; 
  }\label{fig:conductivity_xy_field_frequency}
\end{figure}
\fi

In Fig. \ref{fig:kerr_field} (a) and (b) we show a typical example of Faraday and Kerr rotation angles calculated in 2D limit for a typical even-layer thin films, 
i.e. N = 4 thin film in AF state. 
From the plots we see there is a maximized optical response around the critical electric field that drives the axion insulator state to a semimetal state, which is around 25 meV/nm for 4-layer thin film, and when the frequencies exceed the band gap of MnBi$_2$Te$_4$ thin film, which is around 50 meV. 

\ifpdf
\begin{figure}[htp!]
  \centering
  \includegraphics[width=0.9 \linewidth ]{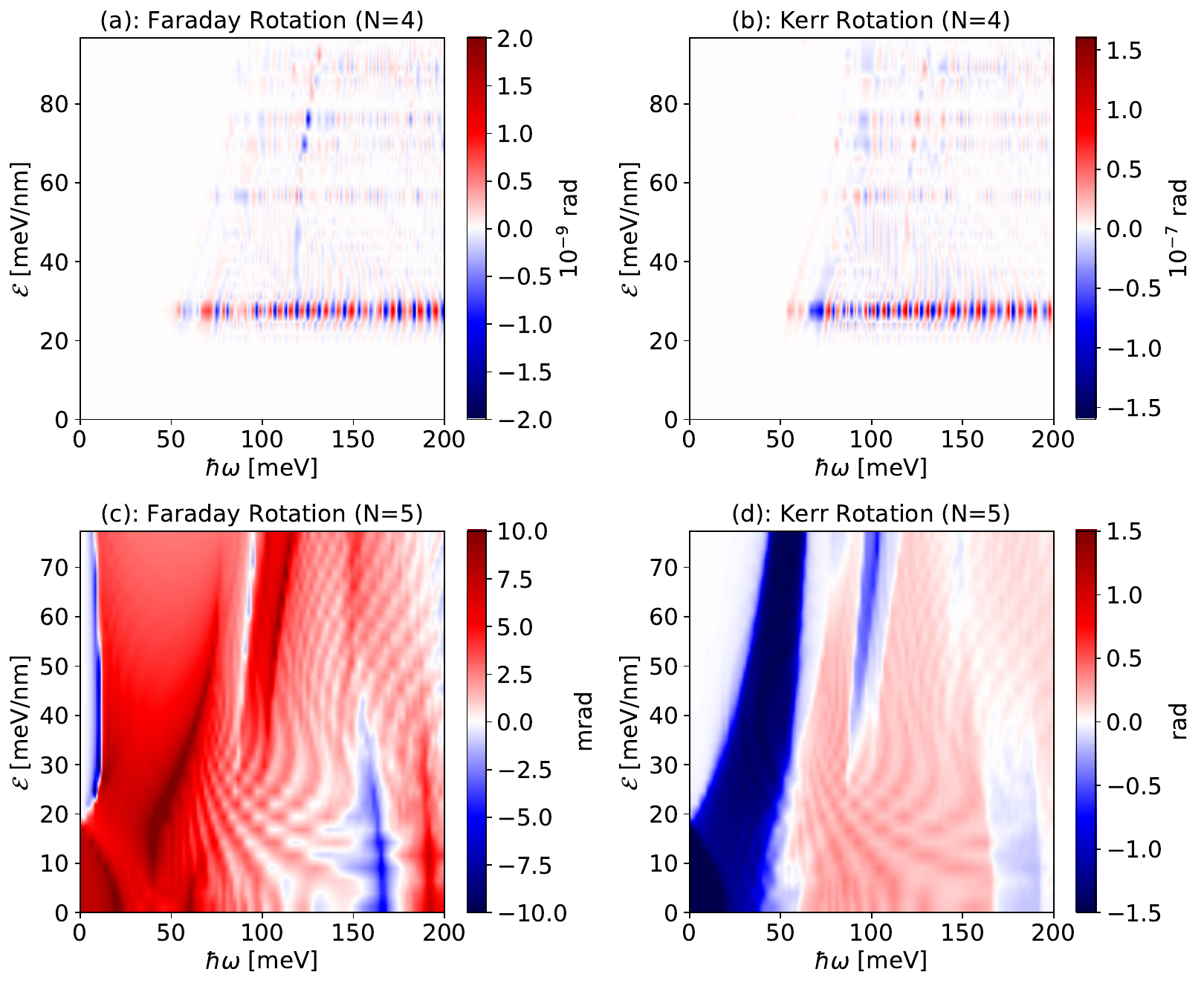}
  \caption{ Electric field dependence of Kerr and Faraday rotation for antiferromagnetic MnBi$_2$Te$_4$ thin films with thickness of 4 and 5 septuple layers.
  (a) and (b) are the plots of Faraday and Kerr rotation angles {\it vs.} optical frequencies and external electric fields for N = 4 MnBi$_2$Te$_4$ thin film;
  (c) and (d) are the plots for  N = 5 MnBi$_2$Te$_4$ thin film.
  }\label{fig:kerr_field}
\end{figure}
\fi

As a comparison, the Faraday and Kerr rotation angles in 2D limit for N = 5 MnBi$_2$Te$_4$ thin film as an example of odd-layer case are shown in Fig. \ref{fig:kerr_field} (c) and (d). We see that the Faraday rotation angles maximize at the DC limit in the absence of electric fields. In the presence of external field the Faraday rotation angles decrease and minimize when the thin film is driven to a semimetal phase, i.e. at the electric field of around 20 meV/nm for 5-layer thin film. Note that this critical field depends on the optical frequencies, that is a positive dependence at low frequencies. Kerr rotation angles have similar dependence on electric field compared with Faraday rotation angles. The Faraday and Kerr rotation signals can thus be used as a detection of the topological phase transition induced by external electric fields.

\bibliography{mbt}

\end{document}